\documentclass[sigconf]{acmart}

\copyrightyear{2026}
\acmYear{2026}
\setcopyright{cc}
\setcctype{by}
\acmConference[WWW '26] {Proceedings of the ACM Web Conference 2026}{April 13--17, 2026}{Dubai, United Arab Emirates.}
\acmBooktitle{Proceedings of the ACM Web Conference 2026 (WWW '26), April 13--17, 2026, Dubai, United Arab Emirates}
\acmDOI{10.1145/3774904.3792123}
\acmISBN{979-8-4007-2307-0/2026/04}

\usepackage{algorithm}
\usepackage{algpseudocode}
\usepackage{multirow}
\usepackage{float}
\usepackage{subfig}
\usepackage{balance} 
\usepackage{hyperref}
\usepackage{url}
\usepackage[inline]{enumitem}

\begin{document}

\title{Iterative Semantic Reasoning from Individual to Group Interests for Generative Recommendation with LLMs}

\settopmatter{authorsperrow=4}
\author{Xiaofei Zhu}
\affiliation{%
  \institution{Chongqing University of Technology}
  \department{College of Computer Science and Engineering}
  \city{Chongqing}
  \country{China}}
\email{zxf@cqut.edu.cn}
\author{Jinfei Chen}
\affiliation{%
  \institution{Chongqing University of Technology}
  \department{College of Computer Science and Engineering}
  \city{Chongqing}
  \country{China}}
\email{htired@stu.cqut.edu.cn}
\author{Feiyang Yuan}
\affiliation{%
  \institution{Chongqing University of Technology}
  \department{College of Computer Science and Engineering}
  \city{Chongqing}
  \country{China}}
\email{pipi77@stu.cqut.edu.cn}
\author{Zhou Yang}
\authornote{Zhou Yang is the Corresponding Author.}
\affiliation{%
  \institution{Chongqing Normal University}
  \department{College of Computer and Information Science}
  \city{Chongqing}
  \country{China}
}
\email{yangzhou@cqnu.edu.cn }


\begin{abstract}
Recommendation systems aim to learn user interests from historical behaviors and deliver relevant items.
Recent methods leverage large language models (LLMs) to construct and integrate semantic representations of users and items for capturing user interests.
However, user behavior theories suggest that truly understanding user interests requires not only semantic integration but also semantic reasoning from explicit individual interests to implicit group interests.
To this end, we propose an Iterative Semantic Reasoning Framework (ISRF) for generative recommendation. ISRF leverages LLMs to bridge explicit individual interests and implicit group interests in three steps. First, we perform multi-step bidirectional reasoning over item attributes to infer semantic item features and build a semantic interaction graph capturing users' explicit interests. 
Second, we generate semantic user features based on the semantic item features and construct a similarity-based user graph to infer the implicit interests of similar user groups. Third, we adopt an iterative batch optimization strategy, where individual explicit interests directly guide the refinement of group implicit interests, while group implicit interests indirectly enhance individual modeling. This iterative process ensures consistent and progressive interest reasoning, enabling more accurate and comprehensive user interest learning.
Extensive experiments on the Sports, Beauty, and Toys datasets demonstrate that ISRF outperforms state-of-the-art baselines.
The code is available at https://github.com/htired/ISRF.
\end{abstract}




\maketitle

\section{Introduction}
 
Recommendation systems~\cite{rendle2010factorizing,autocf2023,Ma2024XRecLL,P5} aim to learn user interests from historical behaviors and recommend the relevant next item.
Early studies~\cite{Tang2018PersonalizedTS,Hidasi2015SessionbasedRW,NGCF} focus on learning sequences of clicked item IDs to capture user interests.
Relying solely on item IDs while ignoring item semantics limits their ability to accurately understand user interests~\cite{ren2024representation,Peng2024DenoisingAW}.

Recent studies incorporate semantic features by leveraging side information (such as brand and price) to better understand user interests.
One line of research~\cite{MoRec,LSMRec,CaFe,ESIF} designs specialized components within small-scale models to integrate semantic features.
MoRec~\cite{MoRec} employs a pre-trained encoder to convert raw item features into embeddings, mitigating the semantic degradation caused by relying solely on ID information.
LSMRec~\cite{LSMRec} uses locality-sensitive hashing to map enhanced item semantic vectors into recommendation representations, enabling semantic integration.
CaFe~\cite{CaFe} employs a coarse-to-fine self-attention framework to fuse user intent with side information.
ESIF~\cite{ESIF} introduces attention and gated fusion mechanisms to jointly update item and side information representations, combined with a denoising module to enhance the utilization of semantic information.
Despite achieving promising results, the limited parameters and knowledge of small-scale models constrain their semantic representation ability~\cite{CIKM23-POD,SLMRec}.

To address this issue, another line of research~\cite{CIKM23-POD,wang2024rdrec,Liu2024LLMEmbLL,SLMRec,wang2024enhancing} introduces large language models (LLMs) with strong representation capabilities.
These methods employ LLMs to encode user and item semantics and combine them with ID features to represent user interests.
LLMEmb~\cite{Liu2024LLMEmbLL} fine-tunes LLMs to generate enriched item semantic representations for capturing user interests.
SLMRec~\cite{SLMRec} distills knowledge from LLM into smaller one, significantly reducing model size while preserving recommendation performance.
POD~\cite{CIKM23-POD} builds upon P5~\cite{P5} by distilling discrete prompts into continuous prompt vectors, thereby enhancing both the expressiveness and efficiency of prompt representations.
ELMRec~\cite{wang2024enhancing} incorporates whole-word embeddings and random feature propagation to enhance semantic representations of users and items, enabling a more precise characterization of user–item interactions and more accurate interest learning.
These methods leverage the strong semantic representation capabilities of LLMs to effectively construct and integrate semantic information, achieving promising results.

\begin{figure}[t] 
  \includegraphics[width=\columnwidth]{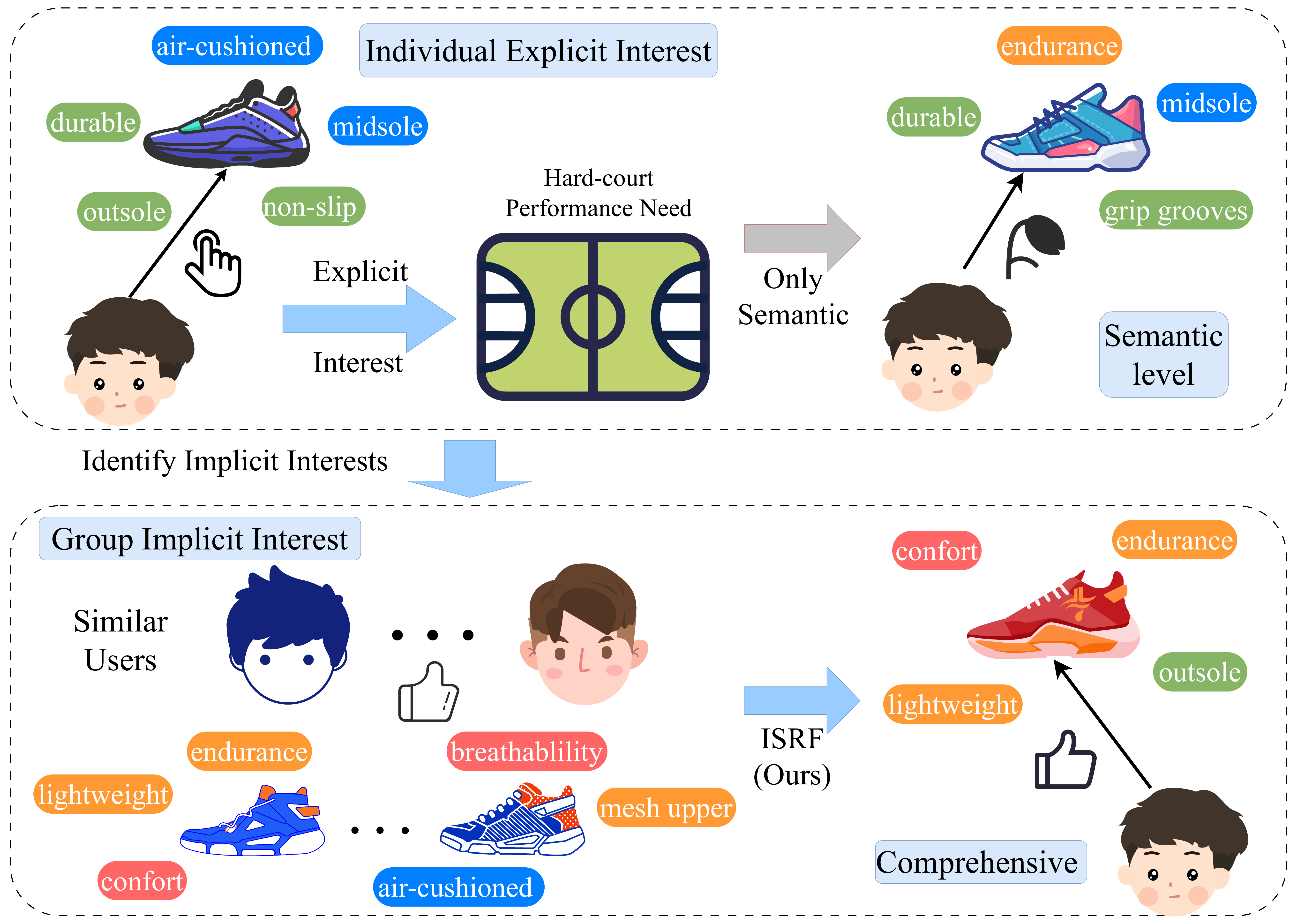}
  \Description{An illustrative comparison between semantic integration approach and our proposed ISRF (Semantic Reasoning). }
  \caption{An illustrative comparison between semantic integration approach and our proposed ISRF (Semantic Reasoning). }
  \label{fig:motivation}
\end{figure}

According to user behavior theories~\cite{gutman1982means,reynolds2001laddering, peter2010consumer}, relying solely on semantic representations is often insufficient to reveal users’ true interests, as uncovering deeper preferences typically requires step-wise reasoning grounded in semantic understanding.
As shown in Figure \ref{fig:motivation}, when a user clicks on a pair of basketball shoes with features such as an air-cushioned midsole and a durable, non-slip outsole, their explicit interest can be inferred as a need for high-impact performance on hard courts.
However, this alone may not reveal the user's deeper interests. Among users with similar behavior, many prioritize breathability and lightweight design during extended games or training sessions to maintain comfort and endurance.
By identifying such group-level implicit interests, the model can recommend more suitable items with greater precision.
This step-wise reasoning process infers explicit interests from item attributes and then uncovers implicit needs through similar user groups, leading to a more comprehensive understanding of user interests.
Nevertheless, effectively simulating this reasoning process to progressively uncover user interests remains a significant challenge.

To address this issue, we propose an \textbf{I}terative \textbf{S}emantic \textbf{R}easoning \textbf{F}ramework (ISRF) for generative recommendation. ISRF leverages LLMs to bridge explicit individual interests and implicit group interests through three coordinated modules:
\begin{enumerate*}[label=(\roman*)]
\item \textbf{Individual Interest Reasoning Module}. We perform multi-step bidirectional reasoning over item attributes to infer semantic item features and build a semantic interaction graph that captures users' explicit interests.
\item \textbf{Group Interest Reasoning Module}. We generate semantic user features from these item features and construct a similarity-based user graph to infer the implicit interests of similar user groups.
\item \textbf{Iterative Refinement Module}. We adopt an iterative batch optimization strategy in which individual explicit interests directly guide the refinement of group implicit interests, and the refined group interests in turn enhance individual preference modeling. This three-stage process ensures consistent, progressive semantic reasoning from explicit to implicit interests, yielding more accurate and comprehensive user-interest representations.
\end{enumerate*}
Extensive results demonstrate that our approach significantly outperforms state-of-the-art baselines. Further analysis shows that each module contributes to stable interest reasoning, leading to the strong overall performance of the proposed framework.

Overall, our main contributions are summarized as follows:

\begin{enumerate}
    \item We introduce a novel semantic reasoning perspective for generative recommendation by inferring explicit individual and implicit group interests, effectively enhancing recommendation performance.
    \item The proposed method performs semantic reasoning from explicit individual interests to implicit group interests and progressively optimizes the reasoning process through an iterative refinement module.
    \item Extensive experiments demonstrate that our model consistently outperforms existing SOTA methods.
\end{enumerate}

\begin{figure*}[t]
  \centering
  \includegraphics[width=\textwidth]{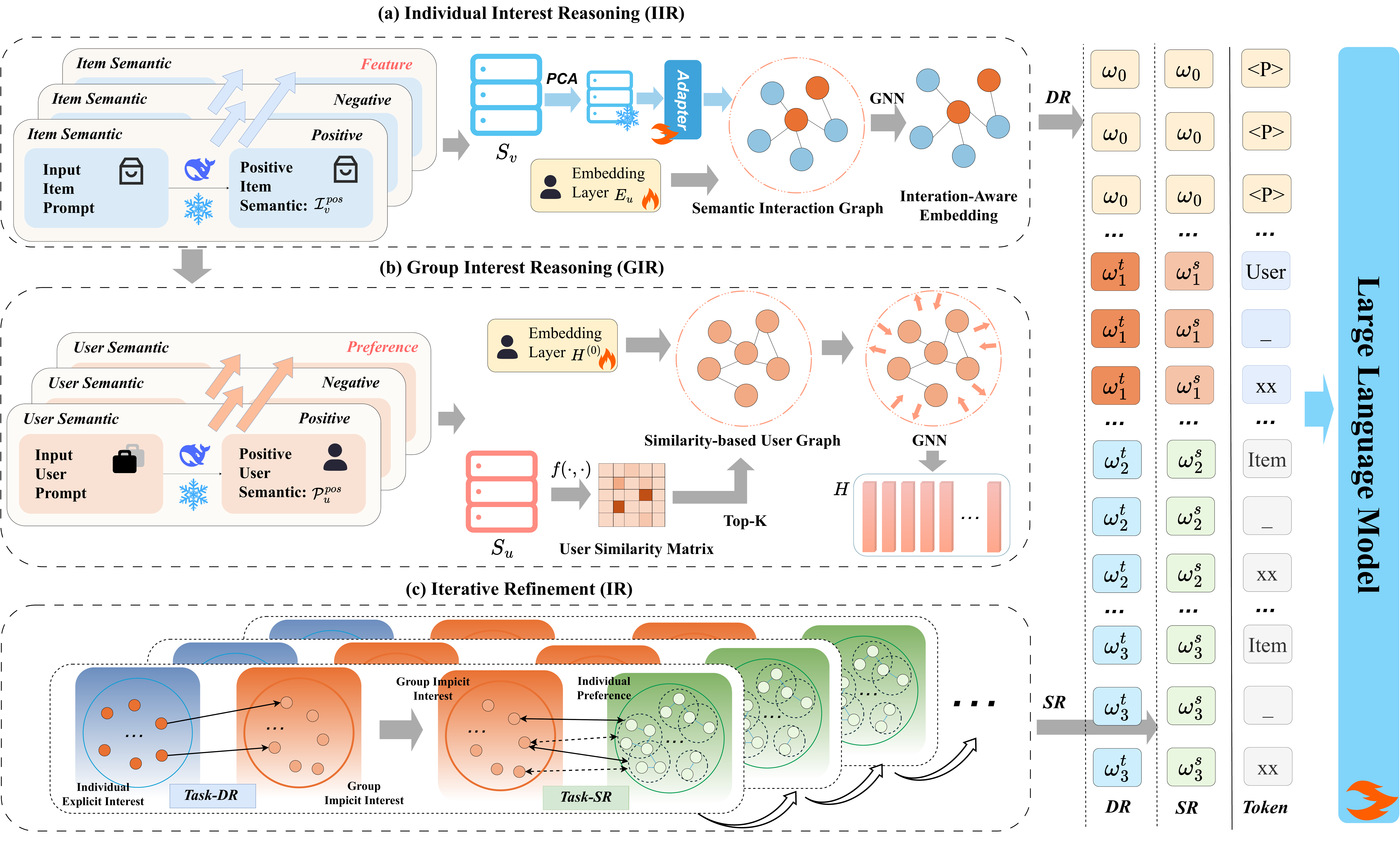}
  \Description{The overall architecture of our proposed Iterative Semantic Reasoning Framework (ISRF)}
  \caption{The overall architecture of our proposed Iterative Semantic Reasoning Framework (ISRF), which includes: (a) Individual Interest Reasoning; (b) Group Interest Reasoning; (c) Iterative Refinement. }
  \label{fig:all}
\end{figure*}

\section{RELATED WORK}

\subsection{Recommendation with Side Information}
Early recommendation methods~\cite{Hidasi2015SessionbasedRW,Tang2018PersonalizedTS,Mao2021SimpleXAS,NGCF} primarily rely on user and item IDs to capture user interests, making it difficult to capture rich semantic information. To address this limitation, researchers have explored side information fusion strategies by incorporating auxiliary attributes such as item titles and categories to enhance recommendation performance. 
MoRec~\cite{MoRec} leverages modality encoders to replace traditional item embeddings.
CaFe~\cite{CaFe} jointly models user intents and item features via coarse-to-fine self-attention.
GCORec~\cite{GCORec} integrates short- and long-term user preferences using various attention mechanisms.
LSMRec~\cite{LSMRec} employs hash-enhanced mapping to improve the effectiveness of pre-trained semantic representations. 
And ESIF~\cite{ESIF} optimizes attention fusion and denoising strategies to better utilize side information and improve accuracy. 
However, due to the limitations of small-scale models in semantic modeling and parameter capacity, these methods still struggle to capture complex user interest patterns effectively.
\subsection{LLMs for Recommendation}

In recent years, large language models (LLMs) have shown great potential in recommender systems (RSs)~\cite{Lin2023HowCR, Wang2024TowardsNL, Bao2024LargeLM, ren2024representation, Tang2024GaVaMoEGG, wang2024rdrec, wang2024enhancing}. A major research direction focuses on enhancing RSs with LLM-generated semantic embeddings~\cite{ren2024representation, Peng2024DenoisingAW, Yang2024DaRecAD, LLM-ESR, Liu2024LLMEmbLL}. 
For example, RLMRec~\cite{ren2024representation} aligns LLM and recommendation models via auxiliary loss, 
and LLM-ESR~\cite{LLM-ESR} combines dual-view modeling with self-distillation to improve performance. 
Recently, generative recommendation has gained attention~\cite{P5, CIKM23-POD, wang2024rdrec, wang2024enhancing}, where LLMs directly generate personalized recommendations. 
POD~\cite{CIKM23-POD} distills discrete prompts into continuous vectors for better prompt representation. 
RDRec~\cite{wang2024rdrec} distills LLM-generated rationales and integrates user-item reviews to enhance reasoning. 
ELMRec~\cite{wang2024enhancing} enhances LLM reasoning via random feature propagation and re-ranking. 
However, these methods primarily focus on shallow semantic modeling and struggle to capture the deep representation and reasoning of user interests.
In contrast, ISRF introduces a novel semantic reasoning perspective and an iterative refinement mechanism to progressively unify user interests.

\section{Problem Definitions}
Following prior research~\citep{wang2024enhancing}, ISRF focuses on two core recommendation tasks: \textit{Sequential Recommendation} and \textit{Direct Recommendation}.
We denote the user and item sets as $\mathcal{U}$ and $\mathcal{V}$, respectively, where $u \in \mathcal{U}$ refers to a user and $v \in \mathcal{V}$ denotes an item.
\setlist[itemize]{left=0em, labelsep=0.5em}

\begin{itemize}
    \item \textbf{Sequential Recommendation}: Given a user $u$ and their historical interaction sequence $\mathcal{V}_u = \{v_1, v_2, \cdots, v_t\}$, the goal is to predict the next item $v_{t+1}$ that the user is likely to interact with.
    \item \textbf{Direct Recommendation}: A candidate set is formed by randomly sampling a positive item $v^+ \in \mathcal{V}_u$ and several negative items $v^- \in \mathcal{V} \setminus \mathcal{V}_u$. The LLM is required to identify the positive item from the candidate set based on user preferences.
\end{itemize}
To handle both tasks within a unified framework, we define the input and output token sequences as
$X \!=\! [x_1, \dots, x_{\lvert X \rvert}]$ and $Y \!=\! [y_1, \dots y_{\lvert Y \rvert}]$, respectively. 
Each input token $x_i$ is associated with an index $z_i$, forming the index sequence $Z \!=\! [z_1, \dots, z_{\lvert X \rvert}]$.  
Let $\mathbf{P}\! =\! [\mathbf{p_1}, \dots, \mathbf{p}_{|\mathbf{P}|}]$ denotes the learnable prompt embeddings, and $\mathbf{X}\! =\! [\mathbf{x_1}, \dots, \mathbf{x}_{|\mathbf{X}|}]$ the embedded input tokens. We concatenate input embeddings with prompts to form the final token input:
$\mathbf{X}_p = [\mathbf{x_1}, \dots, \mathbf{x}_{|\mathbf{X}|},\mathbf{p_1}, \dots, \mathbf{p}_{|\mathbf{P}|}]$. The final input fed into the LLM is computed as:
$\tilde{\mathbf{X}} = \mathbf{X}_p + \beta \mathbf{X}_\omega(Z)$,
where $\beta$ is a scaling factor that controls the contribution of the $\mathbf{X}_\omega(Z)$, and $\mathbf{X}_\omega$ is defined as:
\begin{equation}
\mathbf{X}_\omega=
\begin{cases}
[\omega_0,\omega^s_1,\dots,\omega^s_{|\mathcal{V}|+1}]& \text{if task is SR} , \\
[\omega_0,\omega^t_1,\dots,\omega^t_{|\mathcal{U}|+|\mathcal{V}|}] & \text{if task is DR,} 
\end{cases}
\end{equation}
where $\omega_0$ is a shared embedding for non-ID tokens (e.g., prompts), $\omega_i$ denotes the whole-word embedding of user or item.
\section{METHODOLOGY}
In this section, we introduce the overall architecture of the proposed ISRF framework, as illustrated in Figure~\ref{fig:all}. ISRF comprises three key components:
(a) Individual Interest Reasoning, which models item semantics through multi-step bidirectional reasoning and captures explicit user interests via a graph neural network (GNN);
(b) Group Interest Reasoning, which constructs user semantic preferences based on LLM-enhanced item representations and captures group-level implicit interests using a user semantic graph;
(c) Iterative Refinement, which employs an iterative batch optimization strategy to unify explicit and implicit interests, thereby enhancing the modeling of individual preferences.
\subsection{Individual Interest Reasoning (IIR)}
\label{IIR}

Previous methods typically rely on item attributes for initial semantic reasoning~\cite{ren2024representation,Liu2024LLMEmbLL} or employ auxiliary tasks to enhance semantic understanding~\cite{P5,CIKM23-POD}. However, these approaches underutilize LLMs' reasoning capabilities, resulting in item semantic representations that inadequately capture the diversity of user interests. To this end, we leverage a Chain-of-Thought (CoT)~\cite{COT} reasoning mechanism that guides LLMs to perform multi-step inference on items, generating more interpretable and user-relevant semantic representations. Concurrently, we model individual explicit interests through user-item interaction graphs to enhance the fidelity of interest representation.

Specifically, we prompt the LLM to perform forward reasoning based on the structured attributes of an item, generating a positive description $\mathcal{I}^{pos}_{se}$, such as “what types of users might prefer this item.” Then, conditioned on $\mathcal{I}^{pos}_{se}$, the LLM performs backward reasoning to generate a negative description $\mathcal{I}^{neg}_{se}$, such as “what types of users might dislike this item.” Finally, we fuse $\mathcal{I}^{pos}_{se}$ and $\mathcal{I}^{neg}_{se}$ to form a more diverse and interpretable semantic description $\mathcal{I}_{se}$, e.g., “what key attributes this item may possess.” This chain-of-thought process enables the LLM to progressively infer and understand item semantics, enhancing the accuracy and completeness of the representation. 
The detailed prompt design for items is presented in Appendix~\ref{prompt_sup}.

During training, directly using the enhanced item semantic features $\mathcal{S}_{v} \in \mathbb{R}^{|V|\times d_{llm}}$, i.e, $\mathcal{S}_{v}= \mathcal{T}_{\text{emb}}(\mathcal{I}_{se})$, where $\mathcal{T}_{\text{emb}}(\cdot)$ denotes a pre-trained text encoder~\cite{ren2024easyrec}, as the initial item embeddings may disrupt the original semantic structure. To this end, we apply Principal Component Analysis (PCA)~\citep{FRS1901LIIIOL} to reduce $\mathbf{S}_v$ to a lower-dimensional representation $\tilde{\mathbf{S}}_v \in \mathbb{R}^{|V| \times d_{m}}$, where $d_{m}$ denotes the intermediate embedding dimension. To maintain semantic consistency, we freeze $\tilde{\mathbf{S}}_v$ during training. Then, we use an adapter to map $\tilde{\mathbf{S}}_v$ into the recommendation space, generating the final item embeddings $\mathbf{E}_v$, as shown below:

\begin{equation}
\mathbf{E}_v = W_2(W_1\tilde{\mathbf{S}}_v+b_1)+b_2,
\label{eq:adapter}
\end{equation}
where  $W_1 \in \mathbb{R}^{\frac{d + d_m}{2} \times d_m}$ and $W_2 \in \mathbb{R}^{d \times \frac{d + d_m}{2}}$ are the weight matrices of the projection layers, and $b_1 \in \mathbb{R}^{\frac{d + d_m}{2} \times 1}$ and $b_2 \in \mathbb{R}^{d \times 1}$ are the corresponding bias terms, where $d$ denotes the final embedding dimension in the recommendation space.

Building upon the generated item embeddings $\mathbf{E}_v$, we further apply LightGCN~\citep{He2020LightGCNSA} on the user-item interaction graph $\mathcal{G}$ to model users' explicit interests. The layer-wise message propagation process is defined as follows:  
\begin{equation}
\mathbf{E}^{l+1} = \mathbf{D}^{-1/2} \mathbf{A} \mathbf{D}^{-1/2} \mathbf{E}^{l}, \quad \mathbf{E}^{0} = \begin{bmatrix} \mathbf{E}_u, \mathbf{E}_v \end{bmatrix}^{\text{T}},
\label{eq:gcn_E}
\end{equation}
where $\mathbf{A} \in \mathbb{R}^{(|\mathcal{U}| + |\mathcal{V}|) \times (|\mathcal{U}| + |\mathcal{V}|)}$ denotes the adjacency matrix of the collaborative graph $\mathcal{G}$, and $\mathbf{D}$ is the corresponding degree matrix. $\mathbf{E}_u$ denotes the randomly initialized user embeddings.

After $L$ layers of propagation, the final embedding is obtained by averaging the outputs across all layers:
\begin{equation}
\mathbf{\tilde{E}} = \frac{1}{L+1} \sum_{l=0}^{L} \mathbf{E}^{l}.
\end{equation}
The final embedding matrix $\mathbf{\mathbf{E}}=[\mathbf{\tilde{E}_u}, \mathbf{\tilde{E}_v}]$ effectively contains users' explicit interests and items' contextual semantics.

To further integrate the enhanced embeddings $\mathbf{\tilde{E}}$ into the LLM, we  replace the whole-word embeddings in prompt construction as follows:
\begin{equation}
\omega^t_i=
\begin{cases}
\mathbf{\tilde{e}}_v & \text{if } \omega^t_i \text{ refers to item } v, \\
\mathbf{\tilde{e}}_u & \text{if } \omega^t_i \text{ refers to user } u ,\\
\omega_0 & \text{otherwise},
\end{cases}
\end{equation}
where $\tilde{e}_u$ denotes the embedding of user $u$ from $\mathbf{\tilde{E}_u}$, and $\tilde{e}_v$ denotes the embedding of item $v$ from $\mathbf{\tilde{E}_v}$.

\subsection{Group Interest Reasoning (GIR)}
\label{GIR}
Solely relying on item semantic features from individual interaction histories inadequately captures latent user interests, as behaviorally similar users often share common preferences~\cite{gutman1982means}. To consider such patterns, we design a group interest reasoning module that constructs semantic graphs of similar users via LLM-inferred interest representations. First, analogous to item semantic enhancement $\mathcal{I}_{se}$, we randomly sample a subset of items from each user's interaction history and guide the LLM to generate positive interest descriptions $\mathcal{P}^{pos}_{se}$ through systematic prompting. We then leverage $\mathcal{P}^{pos}_{se}$ as contextual prompts to infer complementary negative interest descriptions $\mathcal{P}^{neg}_{se}$. The final user interest description $\mathcal{P}_{se}$ integrates both perspectives for enhanced interpretability and completeness. Details of the prompt design for users are provided in Appendix~\ref{prompt_sup}.

While semantic representations alone prove insufficient for revealing users' authentic preferences according to user behavioral theory~\cite{gutman1982means, reynolds2001laddering,peter2010consumer}, we propose to model latent preferences of behaviorally similar user groups from a semantic graph perspective. Specifically, we construct a user relation graph based on LLM-enhanced semantic embeddings $\mathbf{S}_u=\mathcal{T}_{emb}(\mathcal{P}_{se})$ , generating a semantic relation matrix $\mathcal{R} \in \mathbb{R}^{|\mathcal{U}| \times |\mathcal{U}|}$. This matrix explicitly captures behavioral similarity among users through semantic information to enhance representational discriminability, with its computation formalized as:

\begin{equation}
\mathcal{R}_{i, j} =
\begin{cases}
1 & \text{if } u_j \in \text{Top-}k(\text{sim}(u_i, \mathcal{U})), \\
0 & \text{otherwise,}
\end{cases}
\label{eq:matrix_p}
\end{equation}
where $\text{Top-}k(\text{sim}(u_i, \mathcal{U}))$ selects the top-$k$ most similar users to user $i$ from the user set $\mathcal{U}$. In this work, we adopt cosine similarity as the similarity metric. By selecting the Top-$k$ most similar users, we reduce computational complexity and improve the efficiency of graph construction.

Subsequently, we utilize a LightGCN on the semantic relation matrix $\mathcal{R}$ to aggregate neighborhood information for refining user representations. The user embedding update process is formulated as:  

\begin{equation} 
\begin{split} 
&\mathbf{H}^{(l)} = \mathbf{D}_{\mathcal{R}}^{-1/2} \mathcal{R} \mathbf{D}_{\mathcal{R}}^{-1/2} \mathbf{H}^{(l-1)},\\
&\mathbf{H} = \frac{1}{L'+1} \sum_{l=0}^{L'} \mathbf{H}^{(l)},
\end{split}
\label{eq:gcn_Hu}
\end{equation}
where $\mathbf{H}$ denotes the final user representation, which incorporates  group implicit interests. $\mathbf{D}_p$ represents the degree matrix of $\mathcal{R}$ to normalize the connections between nodes, and $\mathbf{H}^{(0)}$ denotes the initial user representations obtained by random initialization.

\subsection{Iterative Refinement (IR)}
Although the IIR and GIR stages capture user interests from different perspectives, modeling them independently may result in inconsistent representations and limit their potential complementarity. To address this issue, we introduce an iterative refinement mechanism that facilitates coordinated integration between the two stages. Specifically, in the direct-to-sequential representation alignment phase, the individual explicit interest representation $\tilde{\mathbf{e}}_u$ (Section~\ref{IIR}) is employed as a supervision signal to guide the optimization of the group implicit interest representation $\mathbf{h}_u$ (Section~\ref{GIR}). Conversely, in the sequential alignment phase, $\mathbf{h}_u$ is leveraged to enhance the modeling of individual preference $\mathbf{e}^p_u$ by maximizing the mutual information between them. This iterative refinement process progressively aligns explicit and implicit interest representations, thereby improving both the consistency and generalizability of user modeling.

\subsubsection{Direct-to-Sequential Representation Alignment}
In direct recommendation task, we utilize $\mathbf{h}_u$ as the teacher mediator to guide the optimization of the sequence-based student mediator $\tilde{\mathbf{e}}_u$, thereby enhancing explicit user interest modeling.  To achieve effective alignment, we employ a contrastive distillation loss function that preserves user discriminability while enhancing representation consistency, defined as:
\begin{equation}
\mathcal{L}_{D \rightarrow S} = 
-\frac{1}{B} \sum_{u\in B}\log \frac{f_{c}(\text{sg}[\mathbf{h}_{u}], \tilde{\mathbf{e}}_u)}{\sum_{u' \in B} f_{c}(\text{sg}[\mathbf{h}_{u'}],\tilde{\mathbf{e}}_{u'})},
\label{eq:CD_loss}
\end{equation}
where $f_{c}(\cdot, \cdot) = \exp(\text{sim}(\cdot, \cdot)/\tau)$ denotes the temperature-scaled cosine similarity.

\subsubsection{Sequential Representation Alignment}
We maximize the mutual information between the $\mathbf{h}_u$ and the user preference $\mathbf{e}^p_u$ to enhance the expressiveness of $\mathbf{e}^p_u$ in interest modeling. To this end, we introduce a contrastive loss to align the embedding spaces of $\mathbf{h}^i_u$ and $\mathbf{e}^i_u$, thereby achieving more consistent and discriminative semantic representations. The objective function is defined as follows:

\begin{equation}
\mathcal{L}_{S} = -\frac{1}{B} \sum_{u\in B} \log \frac{f_{c}(\mathbf{h}_u, \mathbf{e}_u)}{\sum_{u' \in B} f_{c}(\mathbf{h}_u, \mathbf{e}_{u'})},
\label{eq:CL_loss}
\end{equation}
where $\mathbf{e}_u$ denotes the interest representation of user $u$, obtained by averaging the full-word embeddings of items in their interaction sequence: $\mathbf{e}_u = \frac{1}{\mathcal{V}_u} \sum_{k=1}^{\mathcal{V}_u} \mathbf{\omega}^s_k$.


\subsection{Optimization and Inference}
In this section, we elaborate on the training and optimization procedures of the ISRF. The corresponding algorithm is presented in Appendix~\ref{Algorithm_sup}.
\subsubsection{Optimization}
To optimize the training process, we adopt a joint loss function that combines the text generation loss and the alignment loss. The overall loss function is defined as:
\begin{equation}
\mathcal{L} =
\begin{cases}
\mathcal{L}_{gen}+\mathcal{L}_{D \rightarrow S} & \text{if task is DR } , \\
\mathcal{L}_{gen}  + \mathcal{L}_{S} & \text{if task is SR},
\end{cases}
\label{eq:loss_all}
\end{equation}
where $\mathcal{L}_{\text{gen}}$ denotes the text generation loss, defined as follows:
\begin{equation}
\mathcal{L}_{\text{gen}} = \frac{1}{|D|} \sum_{(X, Y) \in D} \frac{1}{|Y|} \sum_{t=1}^{|Y|} -\log p(y_t | Y_{<t}, X),
\label{eq:gen_loss}
\end{equation}
where $D$ represents the training dataset containing all input-output pairs, $\lvert D \rvert$ is the total number of samples.
These two components work together to jointly optimize the model parameters.

\subsubsection{Inference}
Following the approach of ELMRec~\cite{wang2024enhancing}, during inference, we employ a beam search algorithm to generate results by selecting the word with the highest likelihood from the vocabulary. This ensures efficient and accurate prediction while maintaining consistency with the training objectives.

\section{EXPERIMENT}

To comprehensively evaluate the effectiveness of the proposed ISRF, we investigate the following six key research questions:

\begin{itemize}
    \item \textbf{RQ1}: How does ISRF perform compared to existing state-of-the-art baselines across different recommendation tasks?
    \item \textbf{RQ2}: What is the impact of individual module designs in ISRF on recommendation performance for distinct tasks?
    \item \textbf{RQ3}: How do different types of semantic information influence recommendation effectiveness?
    \item \textbf{RQ4}: How do key hyperparameters affect the recommendation performance of ISRF?
    \item \textbf{RQ5}: How efficient is ISRF in terms of computational complexity?
    \item \textbf{RQ6}: Does ISRF demonstrate the capability to identify users' implicit interests?
\end{itemize}

\begin{table*}[t]
\centering

\renewcommand{\arraystretch}{1.1}
\resizebox{\textwidth}{!}{
\begin{tabular}{c|cccc|cccc|cccc}
\hline

\textbf{Models} & \multicolumn{4}{c|}{\textbf{Sports}} & \multicolumn{4}{c|}{\textbf{Beauty}} & \multicolumn{4}{c}{\textbf{Toys}} \\
\hline
\textbf{Metrics} & H@5 & N@5 & H@10 & N@10 & H@5 & N@5 & H@10 & N@10 & H@5 & N@5 & H@10 & N@10 \\
\hline
Caser    & 0.0116 & 0.0072 & 0.0194 & 0.0097 & 0.0131 & 0.0087 & 0.0176 & 0.0101 & 0.0166 & 0.0107 & 0.0270 & 0.0141 \\
GRU4Rec  & 0.0129 & 0.0086 & 0.0204 & 0.0099 & 0.0200 & 0.0283 & 0.0137 & 0.0200 & 0.0099 & 0.0059 & 0.0176 & 0.0084 \\
HGN      & 0.0189 & 0.0120 & 0.0313 & 0.0163 & 0.0512 & 0.0266 & 0.0263 & 0.0455 & 0.0201 & 0.0141 & 0.0170 & 0.0300 \\
SASRec   & 0.0233 & 0.0154 & 0.0350 & 0.0192 & 0.0500 & 0.0347 & 0.0170 & 0.0650 & 0.0463 & 0.0306 & 0.0675 & 0.0374 \\
BERT4Rec & 0.0115 & 0.0075 & 0.0191 & 0.0099 & 0.0203 & 0.0124 & 0.0347 & 0.0170 & 0.0116 & 0.0071 & 0.0203 & 0.0099 \\
FDSA     & 0.0182 & 0.0122 & 0.0288 & 0.0156 & 0.0267 & 0.0163 & 0.0407 & 0.0208 & 0.0228 & 0.0140 & 0.0381 & 0.0189 \\
P5       & 0.0387 & 0.0312 & 0.0460 & 0.0336 & 0.0508 & 0.0379 & 0.0644 & 0.0429 & 0.0648 & 0.0567 & 0.0709 & 0.0587 \\
RSL      & 0.0392 & 0.0330 & 0.0512 & 0.0375 & 0.0508 & 0.0381 & 0.0667 & 0.0446 & 0.0676 & 0.0583 & 0.0712 & 0.0596 \\
POD      & 0.0497 & 0.0399 & 0.0585 & 0.0422 & 0.0559 & 0.0420 & 0.0696 & 0.0471 & 0.0692 & 0.0589 & 0.0744 & 0.0601 \\
ELMRec   & \underline{0.0538} & \underline{0.0453} & \underline{0.0616} & \underline{0.0471} & \underline{0.0609} & \underline{0.0486} & \underline{0.0750} & \underline{0.0529} & \underline{0.0713} & \underline{0.0608} & \underline{0.0764} & \underline{0.0618} \\
\textbf{Ours} & \textbf{0.0564} & \textbf{0.0468} & \textbf{0.0648} &\textbf{0.0493} & \textbf{0.0658} & \textbf{0.0526} & \textbf{0.0800} & \textbf{0.0571} & \textbf{0.0741} & \textbf{0.0641} & \textbf{0.0792} & \textbf{0.0652}\\
\hline
Improvement.  
& 4.88$\%^*$ & 3.38$\%^*$ & 5.23$\%^*$ & 4.73$\%^*$ 
& 8.11$\%^*$ & 8.31$\%^*$ & 6.60$\%^*$ & 7.92$\%^*$ 
& 3.92$\%^*$ & 5.44$\%^*$ & 3.68$\%^*$ & 5.54$\%^*$ \\
\hline
\end{tabular}
}
\caption{Performance comparison on the sequential recommendation task, where ``*'' indicates that the improvement is statistically significant ($p$-value $< 0.05$) under a 5-trial $t$-test.}
\label{table:seq_rec_comparison}
\end{table*}

\begin{table*}[htbp]
\centering

\renewcommand{\arraystretch}{1.1}
\resizebox{\textwidth}{!}{
\begin{tabular}{c|cccc|cccc|cccc}
\hline
\textbf{Models} & \multicolumn{4}{c|}{\textbf{Sports}} & \multicolumn{4}{c|}{\textbf{Beauty}} & \multicolumn{4}{c}{\textbf{Toys}} \\
\hline
\textbf{Metrics} & H@5 & N@5 & H@10 & N@10 & H@5 & N@5 & H@10 & N@10 & H@5 & N@5 & H@10 & N@10 \\
\hline
SampleX & 0.2362 & 0.1505 & 0.3290 & 0.1800 & 0.2247 & 0.1441 & 0.3090 & 0.1711 & 0.1958 & 0.1244 & 0.2662 & 0.1469 \\
LightGCN & 0.4150 & 0.3002 & 0.5436 & 0.3418 & 0.4205 & 0.3067 & 0.5383 & 0.3451 & 0.3879 & 0.2874 & 0.5106 & 0.3272\\
NCL & 0.4292 & 0.3131 & 0.5592 & 0.3551 & 0.4378 & 0.3228 & 0.5542 & 0.3607 & 0.3975 & 0.2925 & 0.5120 & 0.3325\\
XSimGCL & 0.3547 & 0.2689 & 0.4486 & 0.2992 & 0.3530 & 0.2734 & 0.4392 & 0.3012 & 0.3351 & 0.2614 & 0.4186 & 0.2885\\
P5 & 0.1955 & 0.1355 & 0.2802 & 0.1627 & 0.1564 & 0.1096 & 0.2300 & 0.1332 & 0.1322 & 0.0889 & 0.2023 & 0.1114 \\
RSL & 0.2092 & 0.1502 & 0.3001 & 0.1703 & 0.1564 & 0.1096 & 0.2300 & 0.1332 & 0.1423 & 0.0825 & 0.1926 & 0.1028 \\
POD & 0.2105 & 0.1539 & 0.2889 & 0.1782 & 0.1931 & 0.1404 & 0.2677 & 0.1639 & 0.1461 & 0.1029 & 0.2119 & 0.1244 \\
ELMRec & \underline{0.5782} & \underline{0.4792} & \underline{0.6479} & \underline{0.4852} & \underline{0.6052} & \underline{0.4852} & \underline{0.6794} & \underline{0.4973} & \underline{0.5178} & \underline{0.4051} & \underline{0.6045} & \underline{0.4141} \\
\textbf{Ours} 
& \textbf{0.6766} & \textbf{0.5535} & \textbf{0.7697} &\textbf{0.5666} & \textbf{0.6773} & \textbf{0.5217} & \textbf{0.7673} & \textbf{0.5352} & \textbf{0.5893} & \textbf{0.4737} & \textbf{0.6733} & \textbf{0.4857}\\
\hline 
Improvement.  
& 23.08$\%^*$ & 20.98$\%^*$ & 24.57$\%^*$ & 22.37$\%^*$ 
& 11.91$\%^*$ & 7.52$\%^*$ & 12.93$\%^*$ & 7.63$\%^*$ 
& 17.01$\%^*$ & 15.51$\%^*$ & 18.80$\%^*$ & 16.78$\%^*$ \\
\hline
\end{tabular}
}
\caption{Performance comparison on direct recommendation task.}
\label{table:topn_rec_comparison}
\end{table*}

\subsection{Experiment Settings}

\subsubsection{Datasets}
In our experiments, we evaluate the proposed method on three widely-used benchmark datasets: Sports \& Outdoors, Beauty, and Toys \& Games\footnote{\url{https://www.amazon.com}}. We adopt the same preprocessing and data splitting protocols as in previous studies~\citep{Zhou2023MMRecSM, wang2024enhancing}. Further details of the datasets are provided in Appendix~\ref{dataset}.

\subsubsection{Baselines}
To evaluate the effectiveness of the proposed ISRF in both direct and sequential recommendation tasks, we compare it with 14 mainstream baselines across four categories.

\begin{enumerate}[wide=0pt, nosep, leftmargin=0pt]
    \item \textbf{Traditional Recommendation Methods}:
    \begin{itemize}
        \item \textbf{SimpleX}~\cite{Mao2021SimpleXAS} enhances representation learning through cosine contrastive loss with large negative sampling. 
        \item \textbf{Caser}~\cite{Tang2018PersonalizedTS} embeds user sequences as pseudo-images and extracts sequential patterns via convolutional operations. 
        \item \textbf{GRU4Rec}~\cite{Hidasi2015SessionbasedRW} replaces traditional item-to-item recommendation by modeling full session sequences. 
        \item \textbf{HGN}~\cite{Ma2019HierarchicalGN} captures users' long- and short-term interests through feature- and instance-level gating mechanisms.
    \end{itemize}

    \item \textbf{Attention-Based Methods}: 
    \begin{itemize}
        \item \textbf{SASRec}~\cite{Kang2018SelfAttentiveSR} models user behavior sequences using self-attention mechanisms. 
        \item \textbf{BERT4Rec}~\cite{Sun2019BERT4RecSR} constructs sequence representations via bidirectional self-attention and masked prediction. 
        \item \textbf{FDSA}~\citep{Zhang2019FeaturelevelDS} jointly models item-level and feature-level sequential patterns to improve recommendation performance.
    \end{itemize}

    \item \textbf{GNN-Based Methods}: 
    \begin{itemize}
        \item \textbf{LightGCN}~\cite{He2020LightGCNSA} streamlines the traditional GCN architecture by removing redundant components, thereby tailoring it specifically for recommendation tasks.
        \item \textbf{NCL}~\cite{Lin2022ImprovingGC} constructs contrastive pairs between users (or items) and their respective structural neighbors to improve the quality of learned embeddings through contrastive learning.
        \item \textbf{XSimGCL}~\cite{Yu2022XSimGCLTE} improves the robustness of user and item representations by generating contrastive views via perturbation with uniform noise.
    \end{itemize}

    \item \textbf{LLM-Based Methods}: 
    \begin{itemize}
        \item \textbf{P5}~\cite{P5} proposes a unified text-to-text paradigm that formulates diverse recommendation tasks as language modeling problems, enabling multi-task generalization and zero-shot prediction through pretraining and personalized prompts.
        \item \textbf{RSL}~\cite{RSL} integrates LLM reasoning with recommendation knowledge for personalized suggestions. 
        \item \textbf{POD}~\cite{CIKM23-POD} enhances recommendation efficiency by distilling discrete prompts into continuous vectors through cyclic training.
        \item \textbf{ELMRec}~\citep{wang2024enhancing} enhances LLMs' recommendation capability by introducing random feature propagation and re-ranking mechanisms.
    \end{itemize}
\end{enumerate}

\subsubsection{Implementation and Metrics}

For semantic reasoning and embedding extraction in user–item interactions, we adopt DeepSeek-R1-14B\footnote{\url{https://ollama.com/library/deepseek-r1:14b}}
 as the backbone large language model to perform multi-step semantic reasoning, and employ EasyRec\footnote{\url{https://huggingface.co/hkuds/easyrec-roberta-large}}~\cite{ren2024easyrec}
 as the semantic embedding extraction module, denoted as $\mathcal{T}_{\mathrm{emb}}$.
Following existing works~\citep{wang2024enhancing}, for direct recommendation, the number of negative items is set to 99 for both training and evaluation. The batch size is set to 64 for training all three tasks. We apply early stopping with a patience of 5 epochs.
P5, POD, ELMRec, and ISRF adopt T5-small~\citep{Raffel2019ExploringTL} as their backbone large language model.
We evaluate all methods using Top-$K$ Hit Rate (H@$K$) and Normalized Discounted Cumulative Gain (NDCG@$K$), where $K \in \{5, 10\}$. All experiments are implemented using the PyTorch framework and conducted on a single NVIDIA GeForce RTX 4090 GPU with 24 GB of VRAM.

\begin{table}[t]
    \centering

    \small 
    \begin{tabular}{ccccccc}
    \toprule
    \multirow{2}{*}{\textbf{Ablation}} & \multicolumn{2}{c}{\textbf{Toys}} & \multicolumn{2}{c}{\textbf{Beauty}} & \multicolumn{2}{c}{\textbf{Sports}} \\
    \cmidrule(lr){2-3} \cmidrule(lr){4-5} \cmidrule(lr){6-7}
    & H@10 & N@10 
    & H@10 & N@10 
    & H@10 & N@10 
    \\
    \midrule
    \multicolumn{7}{c}{\textbf{Sequential Recommendation}} \\
    \textbf{ISRF}
    & \textbf{0.0792} & \textbf{0.0652}
    & \textbf{0.0800} & \textbf{0.0571}
    & \textbf{0.0639} & \textbf{0.0488}
    \\
    \midrule
    w/o $\mathcal{L}_{D \rightarrow S}$
    & {0.0779} & 0.0636 
    & 0.0779 & 0.0553
    & 0.0601 & 0.0464
    \\
    w/o $\mathcal{L}_{S}$
    & 0.0775 & 0.0636 
    & 0.0771 & 0.0548
    & {0.0614} & {0.0469}
    \\
    \midrule
    \multicolumn{7}{c}{\textbf{Direct Recommendation}} \\
    \textbf{ISRF}
    & \textbf{0.6733} & \textbf{0.4857}
    & \textbf{0.7673} & \textbf{0.5352}
    & \textbf{0.7746} & \textbf{0.5768}
    \\
    \midrule
    w/o $\mathcal{I}_{se}$
    & 0.5093 & 0.4248 
    & 0.7209 & 0.5231
    & 0.7170 & 0.5418
    \\
    w/o Adapter 
    & 0.6314 & 0.4592 
    & 0.6868 & 0.5114
    & 0.7097 & 0.5426
    \\
    \bottomrule
    \end{tabular}
    \caption{Ablation studies on Direct Recommendation and Sequential Recommendation tasks across different components of ISRF, evaluated using Hit Rate@10 (H@10) and NDCG@10 (N@10).}
    \label{table:ablation}
\end{table}

\begin{figure}[t]
    \centering
    \subfloat[Direct Recommendation]{
        \includegraphics[width=0.23\textwidth]{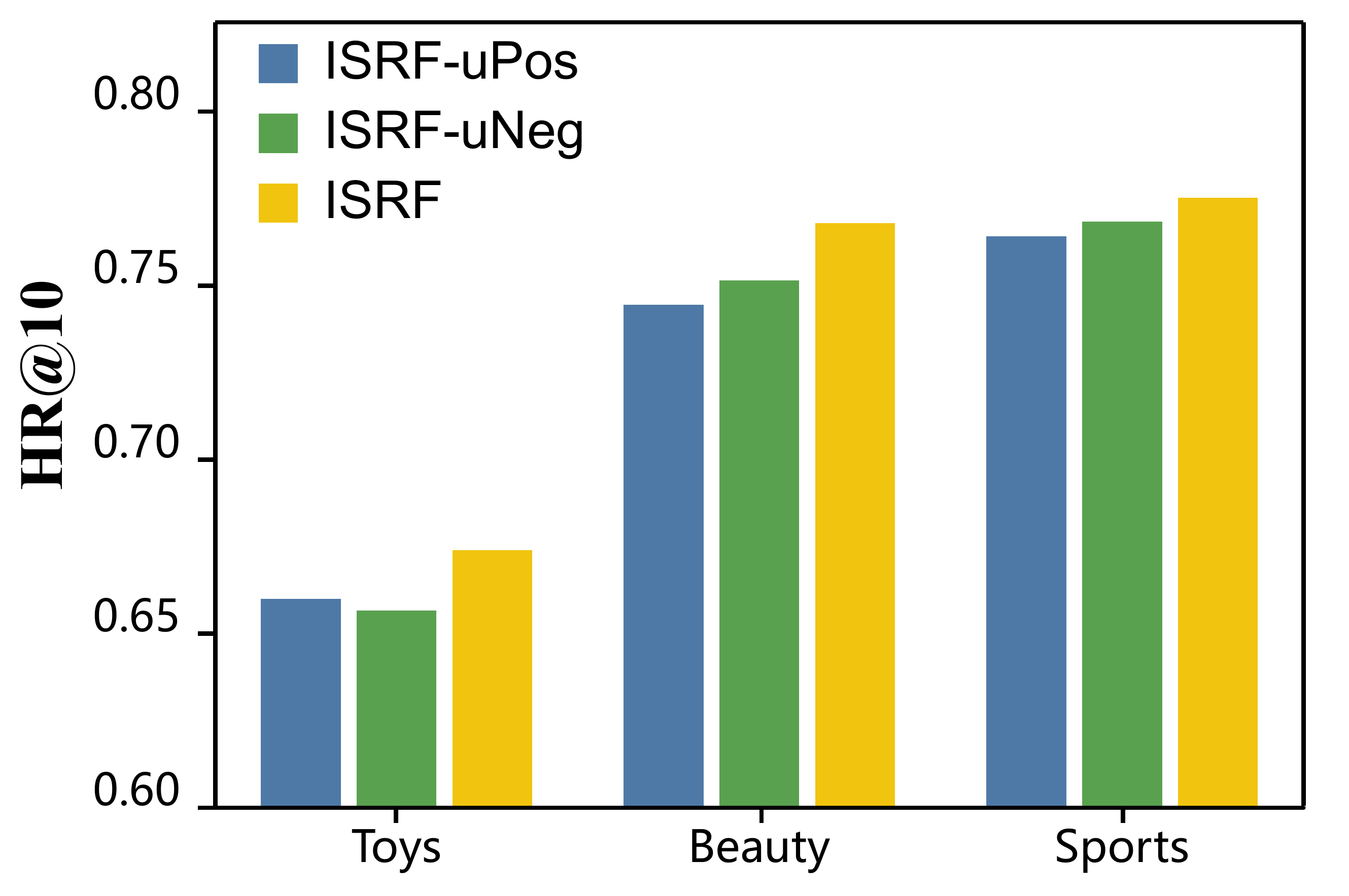}
    }
    \subfloat[Sequential Recommendation]{
        \includegraphics[width=0.23\textwidth]{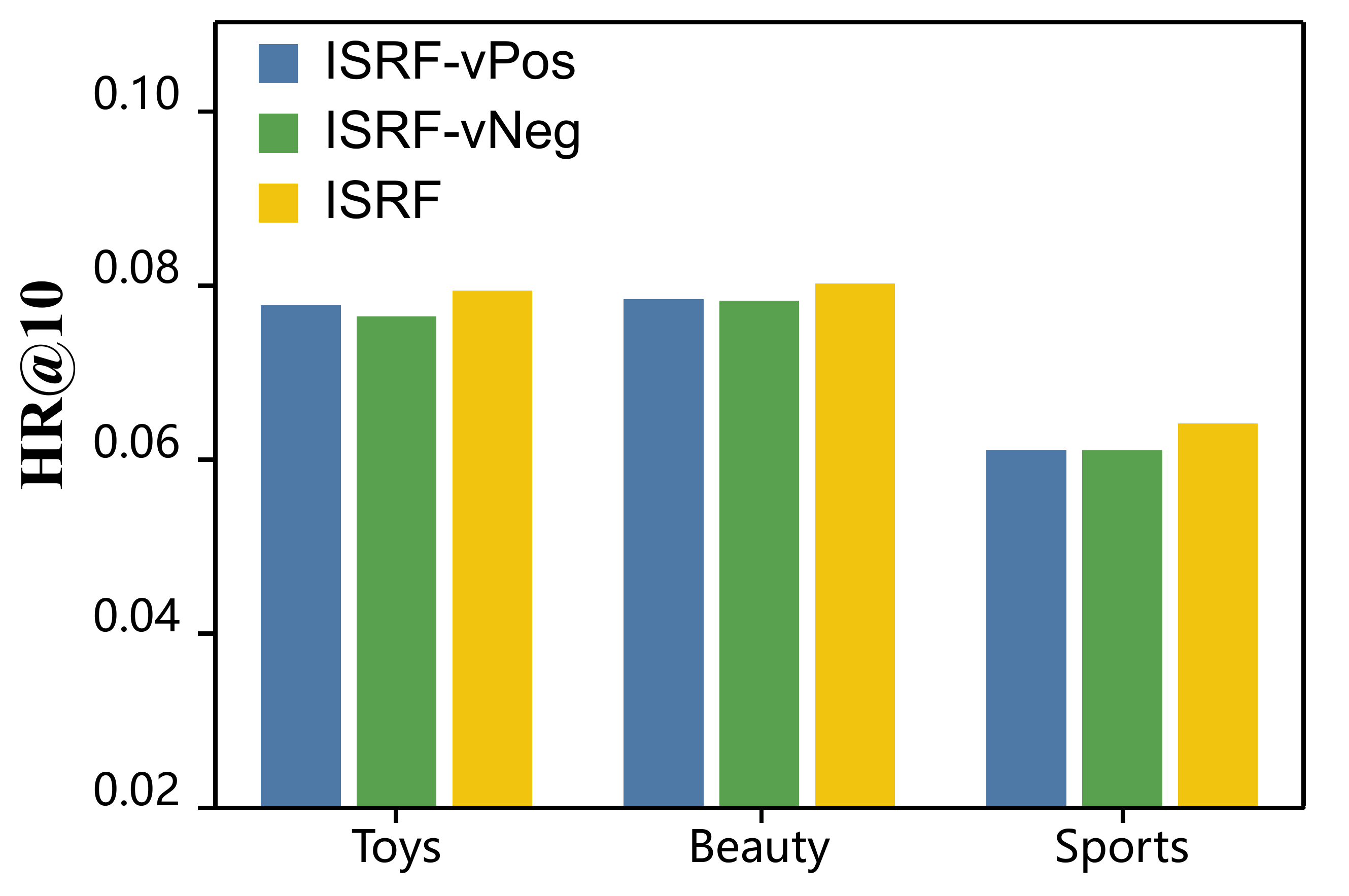}
    }
    \Description{The performance of ISRF and its semantic variants.}
    \caption{The performance of ISRF and its semantic variants.}
    \label{fig:contrast}
\end{figure}

\subsection{Overall Performance (RQ1)}

To validate the effectiveness of the proposed ISRF model, we report its performance on sequential and direct recommendation tasks in Tables~\ref{table:seq_rec_comparison} and~\ref{table:topn_rec_comparison}. 
\begin{itemize}
    \item \textbf{Sequential recommendation}: ISRF also consistently surpasses all baselines on all datasets, with improvements of 3.71$\%$ to 10.37$\%$ over ELMRec. These gains can be attributed to the group interest reasoning module for modeling implicit user interests, as well as the Iterative Refinement mechanism, which effectively optimizes user representations across different granularities.
    \item \textbf{Direct recommendation}: ISRF consistently outperforms all baselines across the three datasets. Compared to the strongest baseline, ELMRec, it achieves performance gains ranging from 7.52$\%$ to 24.57$\%$, primarily due to the individual interest reasoning module’s ability to capture item semantics and fine-grained explicit interests. Moreover, ISRF shows stronger performance on direct recommendation tasks, likely because they rely more heavily on understanding unseen items, highlighting the importance of semantic modeling.
\end{itemize}
Overall, ISRF demonstrates strong generalization and significant performance gains in both sequential and direct recommendation tasks by jointly modeling explicit and implicit interests with iterative refinement.

\begin{figure}[t]
    \centering
    \subfloat[Direct Recommendation]{
        \includegraphics[width=0.23\textwidth]{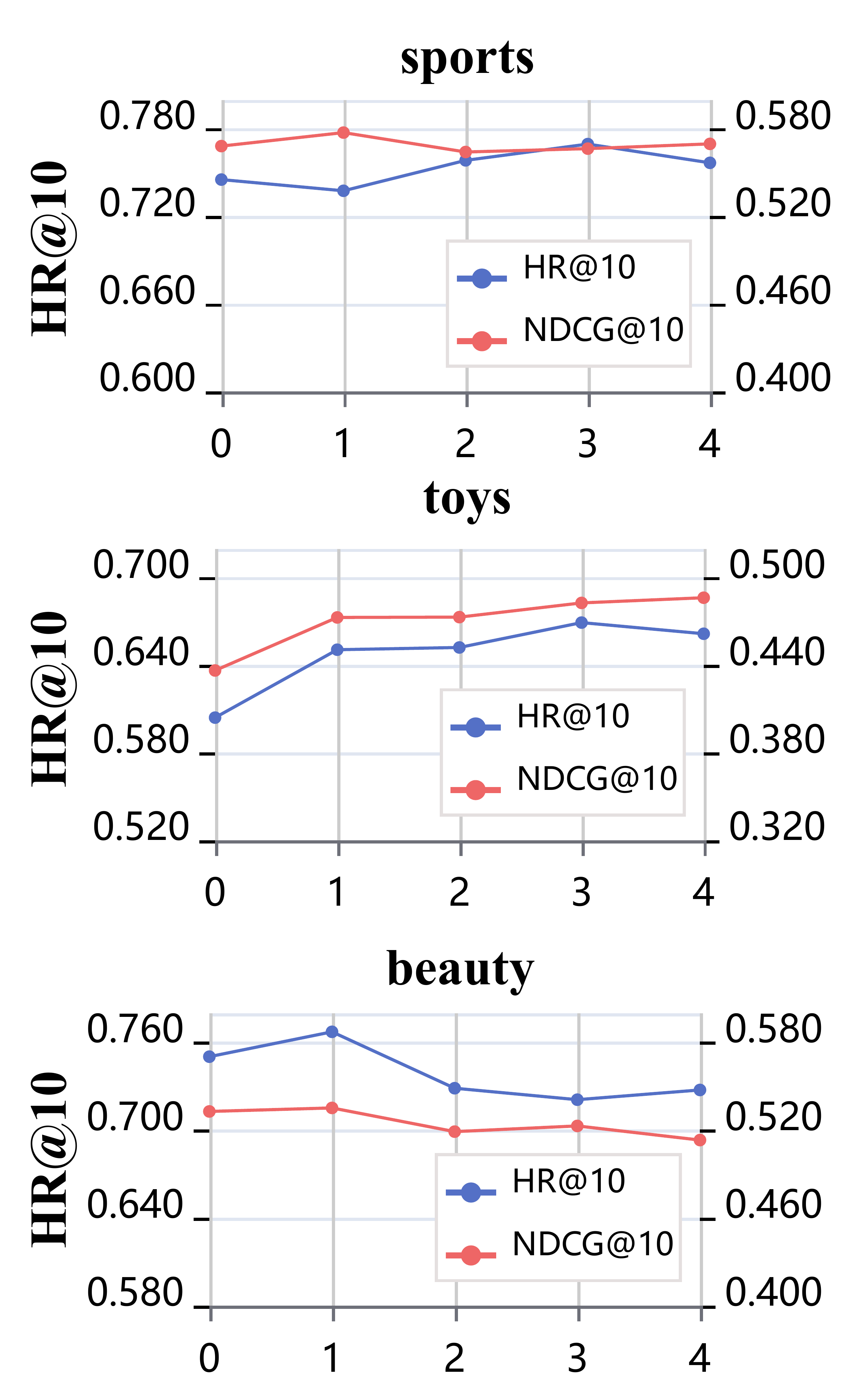}
    }
    \subfloat[Sequential Recommendation]{
        \includegraphics[width=0.235\textwidth]{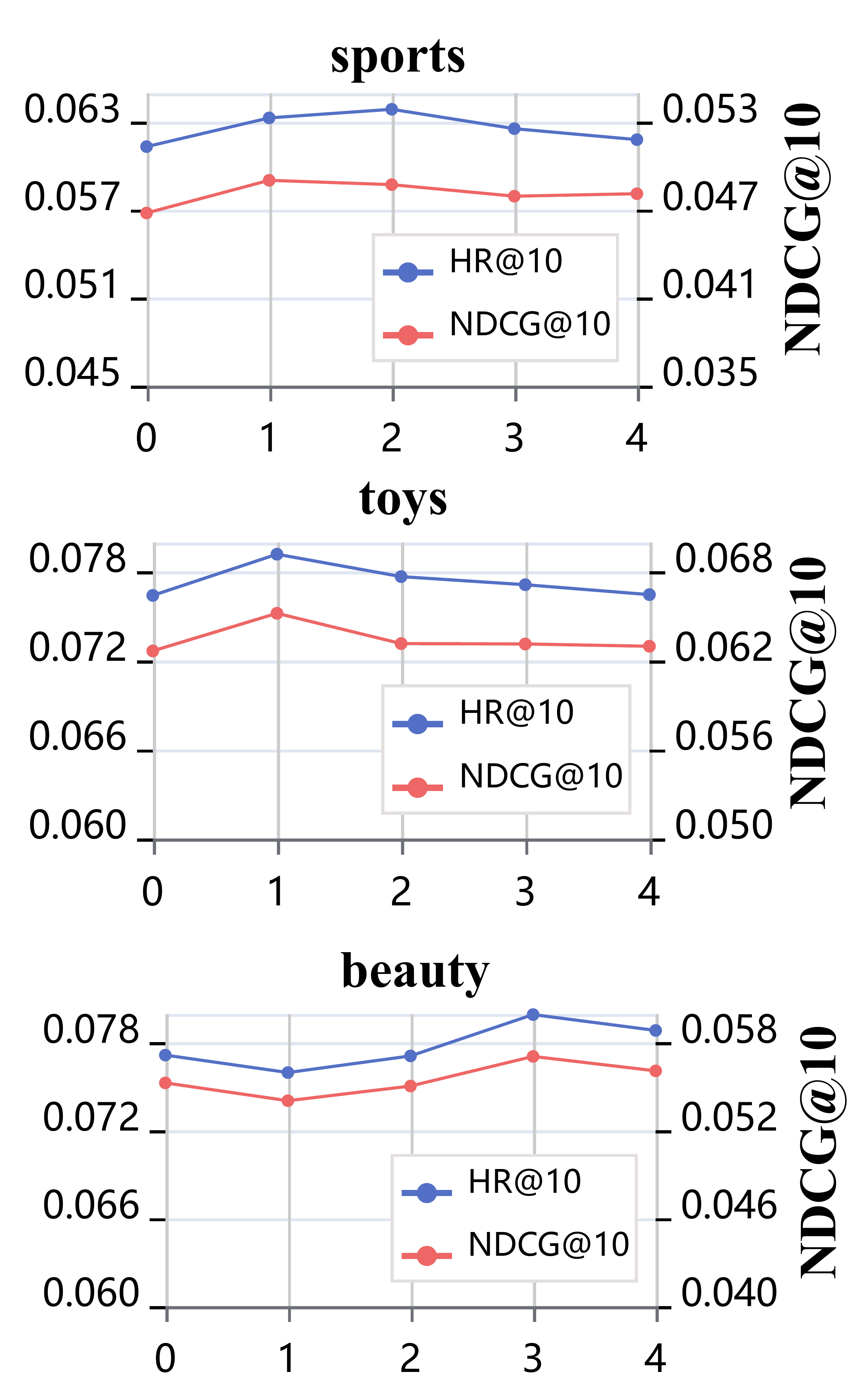}
    }
    \Description{The hyper-parameter study focuses on the $L'$.}
    \caption{The hyper-parameter study focuses on the $L'$.}
    \label{fig:hyper-parameter-L'}
\end{figure}
\subsection{Ablation study (RQ2)}
To further validate the effectiveness of key components in ISRF, we conduct ablation studies on two recommendation tasks, with results shown in Table~\ref{table:ablation}.
We compare the following variants:
\begin{itemize}
    \item \textbf{${w/o}~\mathcal{L}_{D\rightarrow S}$}: Removes the contrastive distillation loss $\mathcal{L}_{D\rightarrow S}$ used to optimize the sequence-based user representation $\tilde{e}_u$ in the direct recommendation task, where temporal preference modeling relies solely on implicit group interests.
    \item \textbf{${w/o}~\mathcal{L}_{S}$}: Removes the contrastive loss, relying solely on user and sequential item ID information for optimization.
    \item \textbf{${w/o}\ \mathcal{I}_{se}$}: Replaces the enhanced item semantic representation $\mathcal{I}_{se}$ with randomly initialized embeddings for the item embedding layer.
    \item \textbf{${w/o}\ Adapter$}: Removes the trainable adapter module, directly using the frozen semantic embeddings $\tilde{\mathbf{S}}_v$.
\end{itemize}

In the sequential recommendation, removing $\mathcal{L}_{D \rightarrow S}$ leads to a notable performance drop, highlighting the importance of explicit interests in guiding group implicit interests learning. Removing $\mathcal{L}_{S}$ also degrades performance, as the LLM then relies only on user and item IDs without semantic enhancement. For direct recommendation, incorporating item semantics $\mathcal{I}_{se}$ improves item understanding, while removing the trainable adapter significantly weakens performance, underscoring its role in aligning semantic and recommendation spaces.

\subsection{Impact of Semantic Variants (RQ3)}

To investigate the impact of different semantic components on performance across tasks, we design several semantic variants of the ISRF inference process:

\begin{itemize}
    \item \textbf{ISRF-uPos}: Uses only the user’s positive semantic reasoning result $\mathcal{P}^{pos}_{se}$ as the final user preference.
    \item \textbf{ISRF-uNeg}: Replaces the final user representation with the negatively inferred user semantics $\mathcal{P}^{neg}_{se}$.
    
    \item \textbf{ISRF-vPos}: Adopts the positively inferred item semantics $\mathcal{I}^{pos}_{se}$ as the item feature.
    \item \textbf{ISRF-vNeg}: Utilizes the negatively inferred item semantics $\mathcal{I}^{neg}_{se}$ to represent the item.
\end{itemize}

The experimental results are illustrated in Figure~\ref{fig:contrast}. In both direct recommendation and sequential recommendation tasks, the full model ISRF consistently outperforms all semantic variants, validating the importance of integrating multi-perspective semantic reasoning encompassing both positive and negative views. Furthermore, we observe that the negative semantic variants (ISRF-uNeg and ISRF-vNeg) exhibit similar performance to the positive ones (ISRF-uPos and ISRF-vPos), suggesting that both positive and negative semantic perspectives make comparable contributions to modeling user interests and act as effective semantic complements.

\subsection{Hyperparameter Sensitivity (RQ4)}

We systematically evaluate the impact of the number of LightGCN layers $L'$ in ISRF. 
As shown in Figure~\ref{fig:hyper-parameter-L'}, increasing $L'$ initially improves performance on both sequential and direct recommendation tasks, 
but the performance plateaus or slightly declines beyond a certain point. 
This suggests that while a moderate propagation depth helps capture implicit user interests, 
excessive layers may lead to over-smoothing or noise accumulation. 
The analysis of the number of top-$K$ similar users $K$ is deferred to Appendix~\ref{app:param_k}.

\begin{table}[t]
\centering
\small 
\resizebox{\linewidth}{!}{
\begin{tabular}{c|c|c|c|c|c}
\toprule
Datsets
& {Models}
& {Train Time}
& {GPU Memory}
& {Infer DR}
& {Infer SR}
\\
\midrule
\multirow{2}{*}{\textbf{Sports}} &
ELMRec
& 10m10s/epoch
& 23.58 GB
& 24m07s
& 13m30s
\\
\cmidrule{2-6}
&
ISRF
& 15m01s/epoch
& 24.19 GB
& 20m23s
& 13m42s
\\
\midrule
\midrule
\multirow{2}{*}{\textbf{Beauty}} &
ELMRec
& 6m13s/epoch
& 21.93 GB
& 12m44s
& 8m19s
\\
\cmidrule{2-6}
&
ISRF
& 8m04s/epoch
& 22.12 GB
& 12m34s
& 8m22s
\\
\midrule
\midrule
\multirow{2}{*}{\textbf{Toys}} &
ELMRec
& 5m07s/epoch
& 21.15 GB
& 9m59s
& 7m29s
\\
\cmidrule{2-6}
&
ISRF
& 5m42s/epoch
& 21.78 GB
& 9m52s
& 7m23s
\\
\bottomrule
\end{tabular}
}
\caption{Computational Cost Comparison. Infer DR and Infer SR denote the inference time of Direct Recommendation and Sequential Recommendation, respectively.}

\label{table:efficiency}
\end{table}

\subsection{Computational Complexity Analysis (RQ5)}
The computational complexity of ISRF mainly stems from the Transformer ($m^2$) and LightGCN ($n^2$), where $m$ denotes the average number of input tokens 
and $n$ represents the total number of user and items. 
Since $m \ll n$, the overall computational complexity of ISRF is dominated by $O(n^2)$, 
which is identical to that of ELMRec. 
As shown in Table~\ref{table:efficiency}, we further report the empirical runtime analysis. 
ISRF consumes computational resources comparable to ELMRec while achieving better performance.

\subsection{Case Study (RQ6)}

\begin{figure}[t]
        \centering
        \includegraphics[width=\columnwidth]{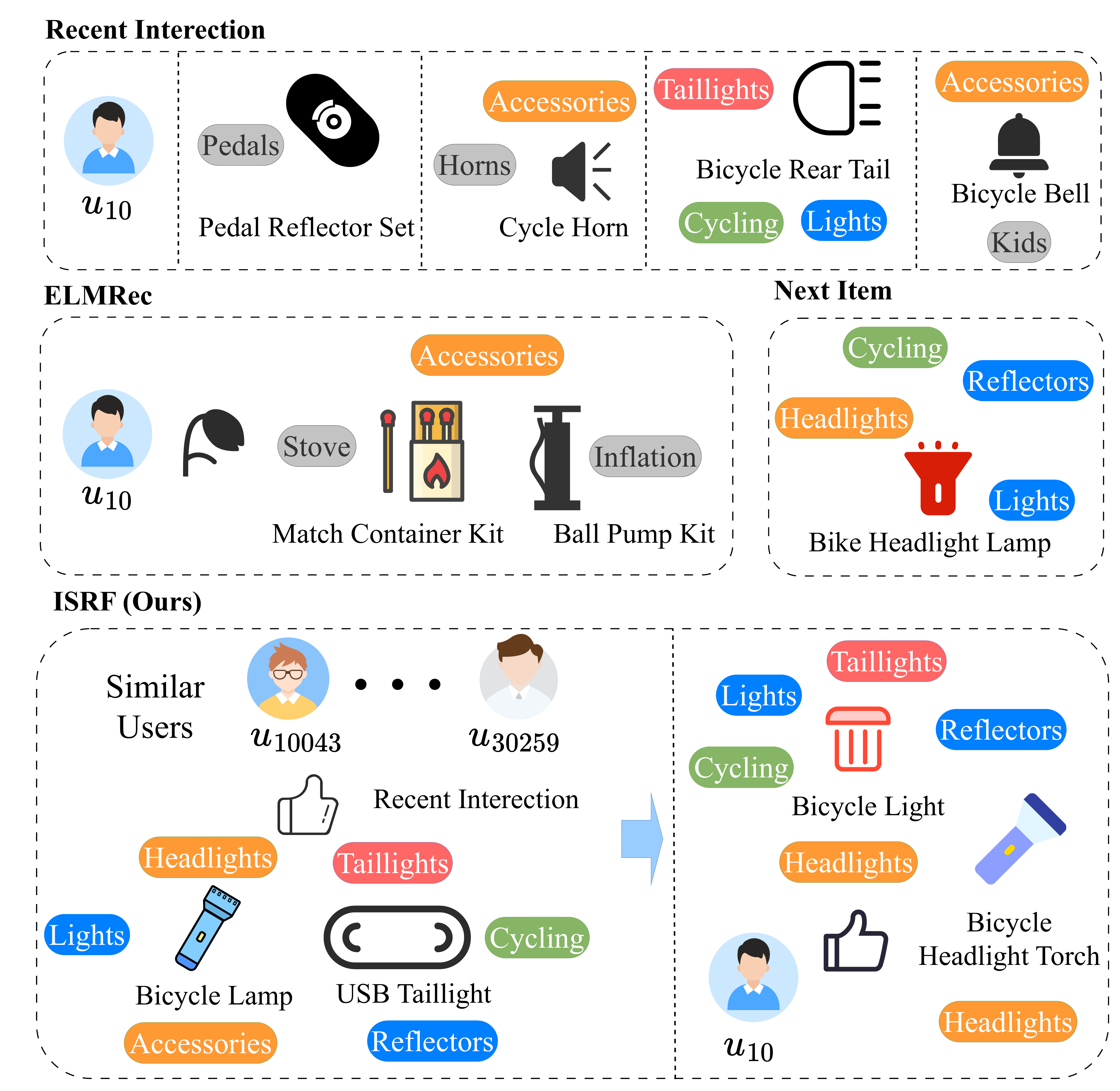}
        \Description{Case study on identifying users’ implicit interests.}
        \caption{Case study on identifying users’ implicit interests.}
        \label{fig:case_study}
\end{figure}

To further verify ISRF’s ability to capture group implicit interests, we present a case study of $u_{10}$ in Figure~\ref{fig:case_study}. While ELMRec mainly recommends items related to the Accessories category (e.g., Match Container Kit, Ball Pump Kit), ISRF identifies the user’s interest in categories like Cycling, Lights, and Headlights. By leveraging the behaviors of semantically similar users (e.g., $u_{10043}$ and $u_{30259}$), who interacted with items in Taillights, Cycling, and Headlights, ISRF infers the user’s implicit Interest for Headlights. As a result, it recommends more relevant items such as Bicycle Light and Bicycle Headlight Torch, demonstrating its advantage in modeling group-level semantic preferences.

\section{Conclusion}
In this paper, we have proposed an Iterative Interest Reasoning Framework (ISRF) for recommendation, which leverages LLMs to perform semantic reasoning from individual explicit interests to group implicit interests by three coordinated modules. First, the individual interest reasoning module infers semantic item features and builds a semantic interaction graph to learn individual explicit interests. Second, the group interest reasoning module constructs a similarity-based user graph to capture the implicit interests of similar user groups. Third, the iterative refinement module alternately optimizes both interests to ensure consistent and progressive reasoning.
Extensive experiments on three real-world datasets demonstrate that ISRF consistently outperforms state-of-the-art baselines.
In future work, we will further enhance the reasoning capabilities of LLMs by incorporating diverse reasoning strategies.

\begin{acks}
This work was supported by the National Natural Science Foundation of China (62472059), the Chongqing Talent Plan Project, China (CSTC2024YCJH-BGZXM0022), the Science and Technology Innovation Key R\&D Program of Chongqing (CSTB2024TIAD-STX0027), the Open Research Fund of Key Laboratory of Cyberspace Big Data Intelligent Security (Chongqing University of Posts and Telecommunications), Ministry of Education (CBDIS202403).
\end{acks}

\bibliographystyle{ACM-Reference-Format}
\bibliography{samples/sample-base}

\appendix

\begin{table}[b]
    \centering
    \begin{tabular}{lccc}
        \toprule
        Dataset & Toys & Beauty & Sports \\
        \midrule
        \#Users & 19,412 & 22,363 & 35,598 \\
        \#Items & 11,924 & 12,101 & 18,357 \\
        \#Reviews & 167,597 & 198,502 & 296,337 \\
        \#Density (\%) & 0.0724 & 0.0734 & 0.0453 \\
        \bottomrule
    \end{tabular}
    \caption{Statistics of the experimental datasets.}
    \label{table:statistic}
\end{table}

\section{SUPPLEMENTARY MATERIAL}
In the supplementary materials, we provide a detailed description of the ISRF algorithmic process and the construction of prompts for both items and users. 
In addition, we include detailed dataset statistics and additional hyperparameter analyses.
\subsection{Prompt Construction}
\label{prompt_sup}
\begin{figure}[t]
  \centering
  \includegraphics[width=\columnwidth]{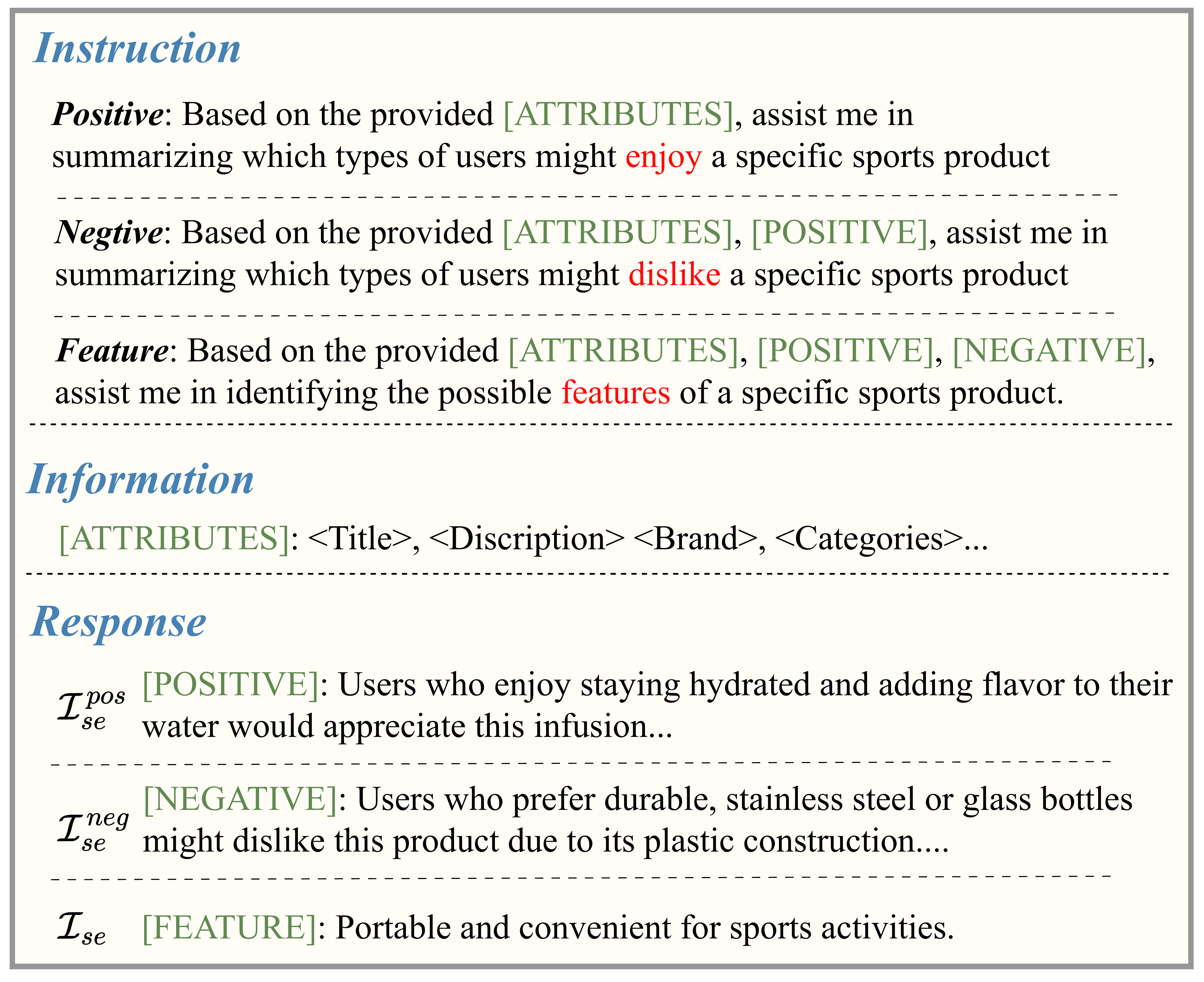}
  \Description{Item Semantic reasoning (Sports)}
  \caption{Item Semantic reasoning (Sports)}
  \label{item_prompt}
\end{figure}

\begin{figure}[t]
  \centering
  \includegraphics[width=\columnwidth]{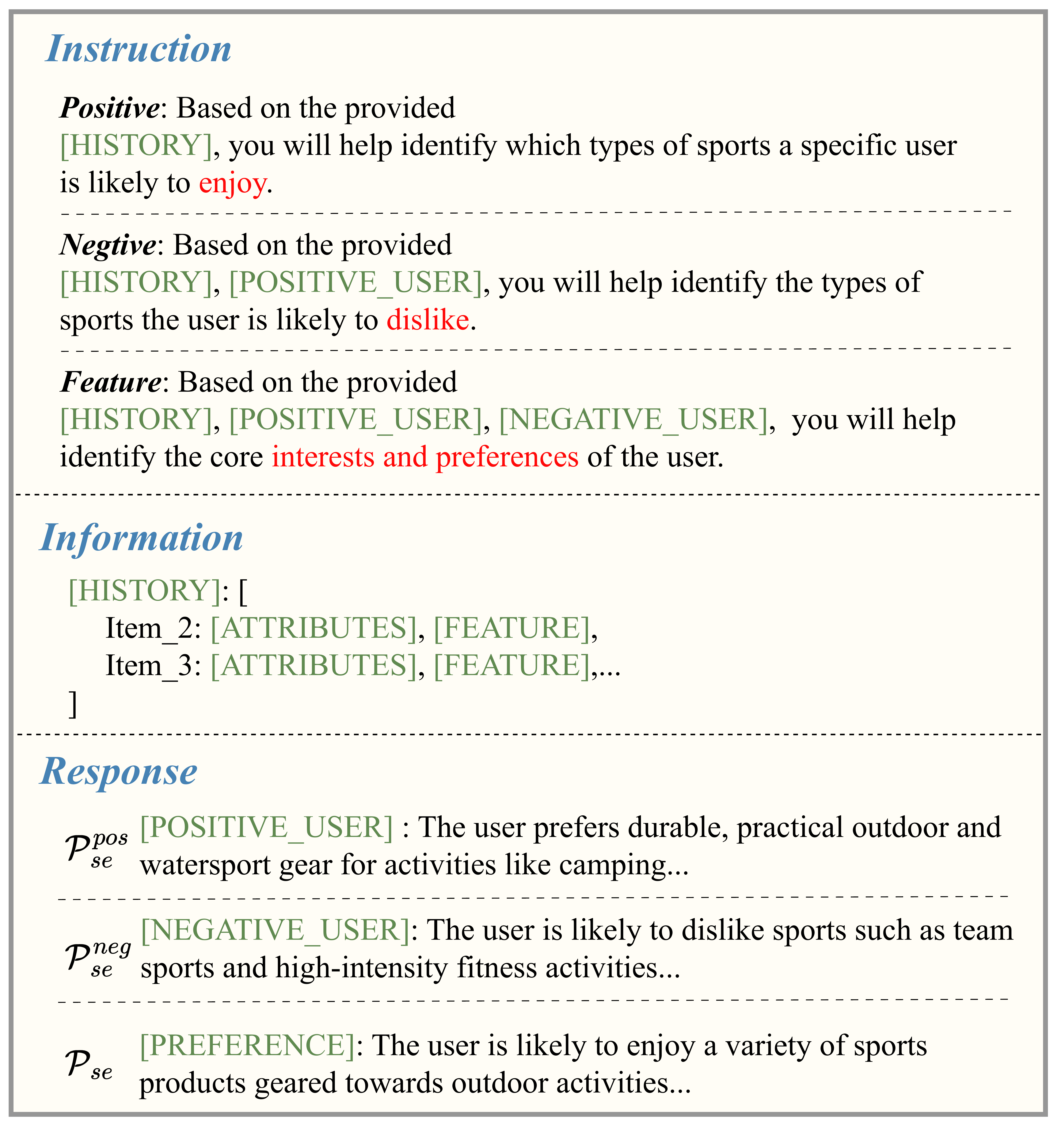}
  \Description{User Semantic reasoning (Sports)}
  \caption{User Semantic reasoning (Sports)}
  \label{user_prompt}
\end{figure}

In this section, we present concrete examples of prompt construction for the Sports dataset. We employ a chain-of-thought strategy to guide the large language model through multi-step reasoning for both items and users, enabling the extraction of comprehensive item features and user preferences, as illustrated in Figure~\ref{item_prompt} and Figure~\ref{user_prompt}.

\subsection{Algorithm for ISRF}
\label{Algorithm_sup}

In this section, we present the algorithmic description of ISRF, as shown in Algorithm~\ref{Algorithm}. 
First, the LLM is initialized as the recommendation model, the intermediate embedding dimension is specified, 
and semantic embeddings for users and items are obtained from the pre-trained LLM (lines 1–3). 
During optimization, the user relation matrix is constructed, item embeddings are reduced via PCA and frozen, 
and user representations are initialized (lines 4–6). 
Then, LightGCN is applied iteratively until convergence (lines 7–8). 
For direct recommendation, item embeddings are mapped into the recommendation space and optimized with user embeddings using contrastive distillation (lines 9–12), 
while for sequential recommendation, a contrastive loss is adopted (lines 13–14). 
In both tasks, the generation loss and total loss are computed to update parameters (lines 16–17). 
Finally, during inference, the trained embeddings and parameters are loaded, and the final recommendation list is generated using beam search (lines 19).

\begin{algorithm}[b]
\caption{Optimization and Inference Process of ISRF}
\begin{algorithmic}[1]
\State Indicate LLM as the recommendation model.
\State Indicate intermediate dimension $d_m$ for LLM-enhanced user and item embeddings.
\State Obtain semantic embeddings $\mathbf{S}_u$ and $\mathbf{S}_v$ for users and items via pre-trained text encoder.
\Statex \textbf{Optimization}
\State Construct the user relation matrix $\mathcal{R}$ by Equation \ref{eq:matrix_p}.
\State Apply PCA to obtain the dimensionality-reduced item semantic embeddings $\tilde{\mathbf{S}}_v$. Freeze $\tilde{\mathbf{S}}_v$.
\State Randomly initialize user embeddings $\mathbf{E}_u,\mathbf{H}^{(0)}$.
\While{not converged}
    \State Using LightGCN, the initial user representations $\mathbf{H}^{(0)}$ are modeled based on the semantic relation matrix $\mathcal{R}$, and the final group
    interest representations $\mathbf{H}$ are optimized through Equation (\ref{eq:gcn_Hu}).
    \If{Task is direct Recommendation}
        \State Map the item semantic representation $\tilde{\mathbf{S}}_v$ to the recommendation space to obtain the item embedding $\mathbf{E}_v$ via Equation (\ref{eq:adapter}).
        \State Use LightGCN to model $\tilde{E}_v$ and $\mathbf{E}_u$ via Equation (\ref{eq:gcn_E}), obtaining the explicit user interest representations $\mathbf{\tilde{E}}_u$ and contextual item semantic representations $\mathbf{\tilde{E}}_v$.
        \State Calculate the contrastive distillation loss $\mathcal{L}_{D \rightarrow S}$ via Equation (\ref{eq:CD_loss}).
    \ElsIf{Task is Sequential Recommendation}
        \State Calculate the contrastive loss $\mathcal{L}_{S}$ via Equation (\ref{eq:CL_loss}).
    \EndIf
    \State Calculate generation loss $\mathcal{L}_{\text{gen}}$ via Equation (\ref{eq:gen_loss}).
    \State Calculate the total loss $\mathcal{L}$ according to Equation (\ref{eq:loss_all}), and update the parameters.
\EndWhile
\Statex \textbf{Inference}
\State Generate final recommendation list using beam search by selecting the word with the highest likelihood from the vocabulary.
\end{algorithmic}
\label{Algorithm}
\end{algorithm}

\subsection{Dataset Details}
\label{dataset}

In this section, we evaluate the proposed method on three widely-used benchmark datasets: 
Sports \& Outdoors, Beauty, and Toys. 
These datasets are collected from Amazon and span different product domains with diverse user–item interaction patterns. 
Following prior works~\citep{Zhou2023MMRecSM,wang2024enhancing}, 
we process item attribute information using the same strategy and adopt the same data splitting protocol. 
Specifically, for direct and sequential recommendation tasks, 
we use the last interaction of each user for testing, 
the second-to-last interaction for validation, 
and the remaining interactions for training.
Detailed statistics of the datasets are provided in Table~\ref{table:statistic}.

\subsection{Hyperparameter Sensitivity}
\label{app:param_k}
\begin{figure}[t]
    \centering
    \subfloat[Direct Recommendation]{
        \includegraphics[width=0.23\textwidth]{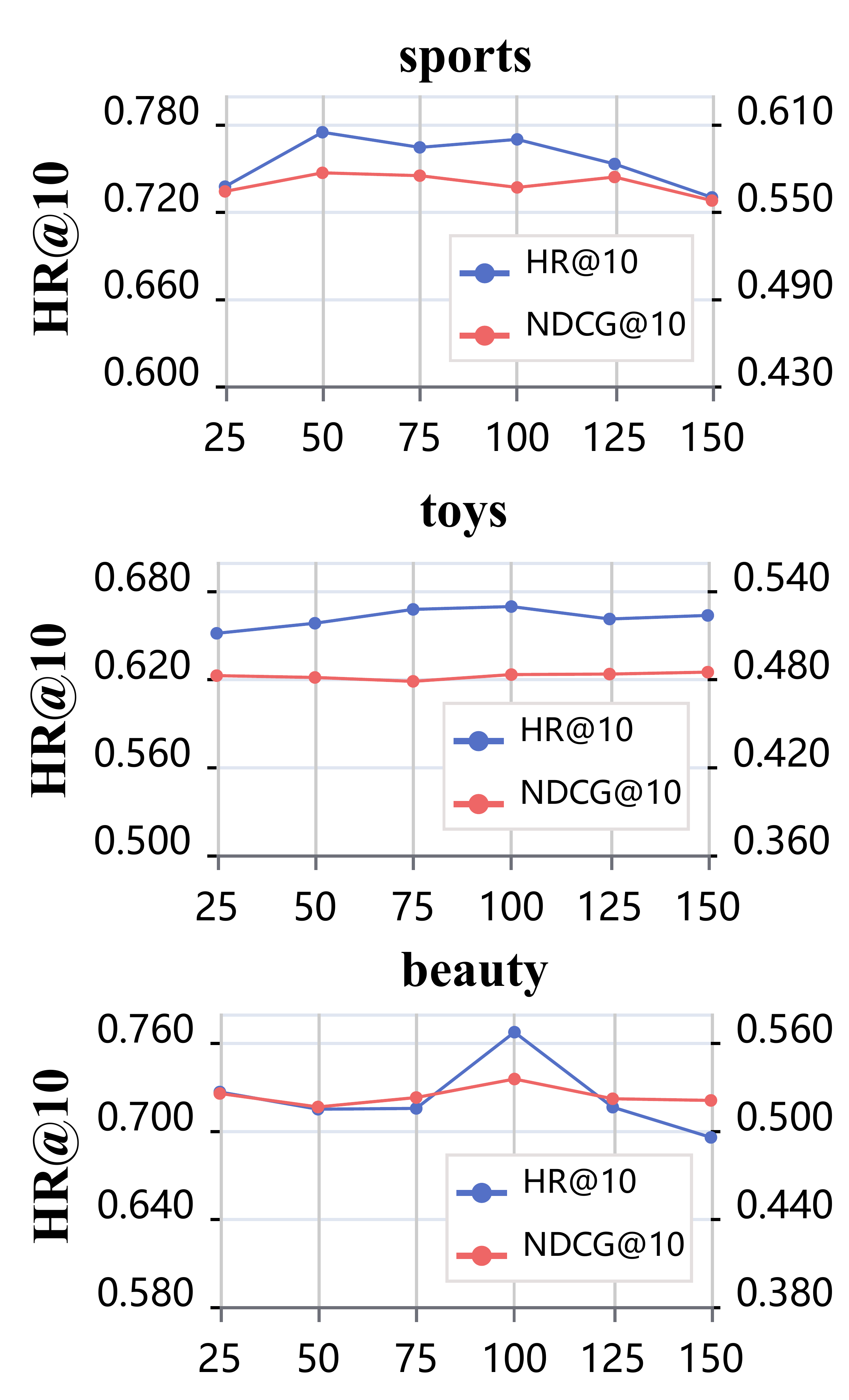}
    }
    \subfloat[Sequential Recommendation]{
        \includegraphics[width=0.23\textwidth]{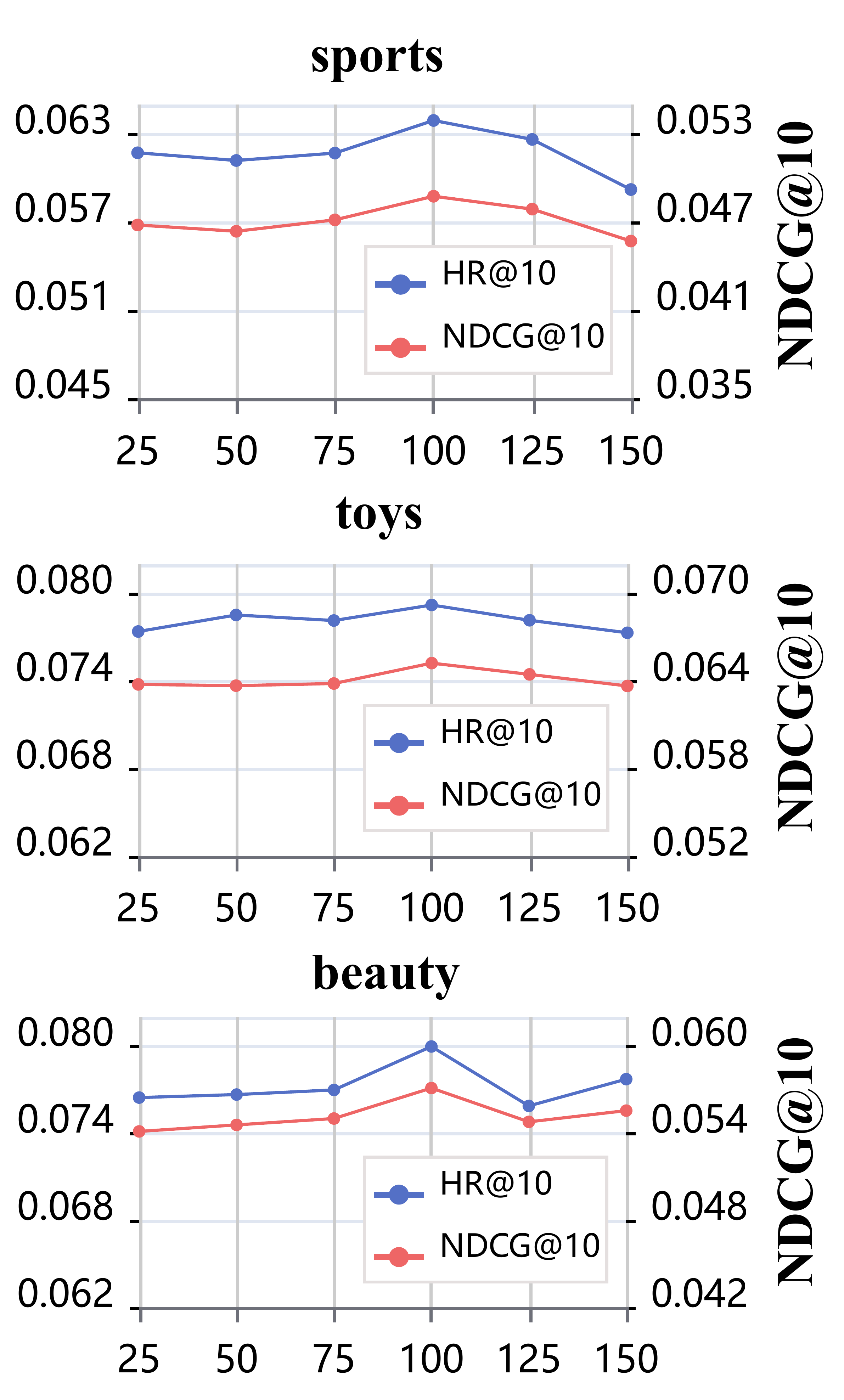}
    }
    \Description{The hyper-parameter study focuses on the $K$.}
    \caption{The hyper-parameter study focuses on the $K$.}
    \label{fig:hyper-parameter-K}
\end{figure}

Figure~\ref{fig:hyper-parameter-K} illustrates the effect of the hyperparameter Top-$K$ similar users. Performance first improves and then degrades as $K$ increases. For sequential recommendation, the best results on all three datasets are achieved at $K=100$. For direct recommendation, the optimal $K$ is 100 on the Beauty and Toys datasets and 50 on the Sports dataset. These results indicate that incorporating a larger set of similar users benefits preference modeling, while an excessively large $K$ may introduce noise and degrade accuracy.

\end{document}